\documentclass[manuscript]{aastex}

\usepackage{lscape}
\usepackage[scriptsize]{caption}
\usepackage{longtable}

\slugcomment{...}

\shorttitle{A multiwavelength study on the high-energy behaviour of Fermi/LAT pulsars}
\shortauthors{Marelli, M. et al.}

\begin{document}

\title{A multiwavelength study on the high-energy behaviour of the Fermi/LAT pulsars}

\author{Martino Marelli\altaffilmark{1,2}}

\author{Andrea De Luca\altaffilmark{1,3}}

\and

\author{Patrizia A. Caraveo\altaffilmark{1}}

\altaffiltext{1}{INAF/IASF Milano - Via E. Bassini 15 - I-20133 Milano - Italy}
\altaffiltext{2}{Universit\'a degli Studi dell'Insubria - Via Ravasi 2 - 21100 Varese - Italy}
\altaffiltext{3}{IUSS - V.le Lungo Ticino Sforza 56 - 27100 Pavia - Italy}

\begin{abstract}
Using archival as well as freshly acquired data, we assess the X-ray behaviour
of the Fermi/LAT $\gamma$-ray pulsars listed in the First Fermi source catalog \citep{Abd10b}.
After revisiting the relationships between the pulsars' rotational energy losses and
their X and $\gamma$-ray luminosities, we focus on the distance-indipendent $\gamma$ to 
X-ray flux ratios.
When plotting our F$_{\gamma}$/F$_X$ values as a function of the pulsars' rotational energy losses, 
one immediately sees that pulsars with similar energetics have F$_{\gamma}$/F$_X$
spanning 3 decades. Such spread, most probably stemming from vastly different
geometrical configurations of the X and $\gamma$-ray emitting regions,
defies any straightforward interpretation of the plot.
Indeed, while energetic pulsars do have low F$_{\gamma}$/F$_X$ values, little can
be said for the bulk of the Fermi neutron stars.
Dividing our pulsar sample into radio-loud and radio-quiet subsamples, we find
that, on average, radio-quiet pulsars do have higher values of F$_{\gamma}$/F$_X$,
implying an intrinsec faintness of  their X-ray emission and/or a different geometrical configuration.
Moreover, despite the large spread mentioned above,
statistical tests show a lower scatter in the radio-quiet dataset with
respect to the radio-loud one, pointing to a somewhat more constrained geometry for the radio-quiet objects with respect to the
radio-loud ones.
\end{abstract}


\section{Introduction} \label{bozomath}

The vast majority of the 1800 rotation-powered pulsars known to date \citep{Man05} were discovered
by radio telescopes. While only few pulsars have also been seen in the optical band \citep[see e.g.][]{Mign08,Mign10}, 
the contribution of Chandra and XMM-Newton telescopes increased the number of
X-ray counterparts of radio pulsars bringing the gran total of to $\sim$100 \citep[see e.g.][]{Bec09}.
Such high-energy emission can yield crucial information on the pulsar physics, 
disentangling thermal components from non-thermal ones, and tracing the presence of pulsar wind nebulae (PWNe).\\
Chandra's exceptional spatial resolution made it possible
to discriminate clearly the PWN and the PSR contributions while XMM-Newton's high spectral
resolution and throughput unveiled the multiple spectral components which characterize pulsars (see e.g. \citet{Pos02}).
Altough the X-ray non-thermal powerlaw index seems somehow related to
the gamma-ray spectrum \citep[see e.g.][]{Kas04}, extrapolating the X-ray data
underpredicts the $\gamma$-ray flux by at least one order of magnitude \citep[see e.g.][]{Abd10d}.\\
Until the launch of Fermi, only seven pulsars were seen in
high-energy gamma rays \citep{Tom08}, and only one of them, Geminga, was not detected by radio telescopes.
The Fermi Large Area Telescope (LAT) changed dramatically such scenario estabilishing radio-quiet pulsars
as a major family of $\gamma$-ray emitting neutron stars. After one year of all-sky monitoring Fermi/LAT has detected 54 gamma-ray
pulsars, 22 of which are radio-quiet \citep{Abd09,Saz10,Cam09}.
Throughout this paper we shall classify as radio-quiet all the pulsars detected by Fermi through blind searches \citep{Abd09,Saz10}
but not seen in radio in spite of dedicated deep searches.
Containing a sizeble fraction of radio-quiet pulsars, the Fermi sample provides,
for the first time, the possibility to compare the phenomenology of radio-loud and radio-quiet neutron stars
assessing their similarities and their differences (if any).\\
While our work rests on the Fermi data analysis and results \citep{Abd09,Saz10} for the X-ray side we had
first to build an homogeneus data set relying both on archival sources and on fresh observations.\\
In the following we will address the relationship between the classical pulsar parameters, such as
age and overall energetics $\dot{E}$, and their X and $\gamma$-ray yields.
While the evolution of the X and $\gamma$-ray luminosities as a function of $\dot{E}$ and the characteristic age
$\tau_c$ have been already discussed, we will concentrate on the ratio between the X and $\gamma$-ray luminosities
thus overcoming the distance conondrum which has hampered the studies discussed so far in the literature. 
We note that F$_{\gamma}$/F$_X$ parameter probes both pulsar efficiencies at different wavelenghts
and distribution of the emitting regions in the pulsar magnetosphere. Thus, such a distance-indipendent
approach does magnify the role of both geometry and geography in determining the high-energy emission from pulsars.

\section{Data Analysis}

\subsection{$\gamma$-ray Analysis}
\label{gdata}

We consider all the pulsars listed in the First Year Catalog of Fermi $\gamma$-ray sources
\citep{Abd10b} which contains the $\gamma$-ray pulsars listed in the First Fermi
pulsar catalog \citep{Abd09} as well as the new blind search pulsars found by \citet{Saz10}.
Our sample comprehends 54 pulsars:\\
- 29 detected using radio ephemerides\\
- 25 found through blind searches; of these 3 were later found to have also a radio emission
and, as such, they were added to the radio emitting ones.\\
Thus, our sample of $\gamma$-ray emitting neutron stars consists of 32 radio pulsars and 22
radio-quiet pulsars.
Here, we summarize the main characteristics of the analysis performed in the two articles.\\
The pulsar spectra were fitted with an exponential cutoff powerlaw model of the form:

\
$dN/dE = KE_{GeV}^{-\gamma}exp(-E/E_{cutoff})$
\

1 GeV has been chosen to define the normalization factor because it's the energy at which the relative uncertainty
on the differential flux is minimal.\\
The spectral analysis was performed taking into account the contribution of all the neighboring sources
(up to 17$^{\circ}$) and the diffuse emission. Sources at more than 3$^{\circ}$ from any pulsars were assigned fixed spectra,
taken from the all-sky analysis.
$\gamma$-rays with E$>$100 MeV have been used and the contamination produced
by cosmic rays interactions in the Earth's atmosphere was avoided by selecting a zenith angle $>$105$^{\circ}$.\\
At first, all events have been used in order to obtain a phase-averaged spectrum for each pulsar.
Next the data have been splitted into on-pulse and off-pulse samples. The off-pulse 
sample has been described with a simple powerlaw while, for the on-pulse emission, an exponentially cutoff powerlaw has
been used, with the off-pulse emission (scaled to the on-pulse phase interval) added to the model.
Such an approach is adopted in order to avoid a possible PWN contamination to the pulsar spectrum.\\
For completeness, we included in our sample also the 4 radio pulsars listed in the 4th IBIS/ISGRI catalog \citep{Bir09} but,
so far, not seen by Fermi. 
Searching in the 1-year Fermi catalog \citep{Abd10b}, we found a potential counterpart for PSR J0540-6919 but the lack
of a pulsation prevent us to associate the IBIS/ISGRI pulsar with the Fermi source. We therefore used the 1FGL flux as an upper limit.
The three remaining IBIS pulsars happen to be located near the galactic centre,
where the intense radiation from the disk of our Galaxy hampers the detection of $\gamma$-ray sources.
We used the sensitivity map taken from \citet{Abd10b} to evaluate the Fermi flux upper limit.

\subsection{X-ray Data}
\label{xdata}

The X-ray coverage of the Fermi LAT pulsars is uneven since the majority of the newly discovered radio-quiet PSRs 
have never been the target of a deep X-ray observation,
while for other well-known $\gamma$-ray pulsars - such as Crab, Vela and Geminga - one can rely on a lot of observations.
To account for such an uneven coverage, we classify the X-ray spectra on the basis of the public X-ray data available, thus assigning:\\
- label "0" to pulsars with no confirmed X-ray counterparts (or without a non-thermal spectral component);\\
- label "1" to pulsars with a confirmed counterpart but too few photons to assess its spectral shape;\\
- label "2" to pulsars with a confirmed counterpart for which the data quality allows for the analysis of both the pulsar and the nebula (if present).\\
An "ad hoc" analysis was performed for seven pulsars for which the standard analysis couldn't be applied (e.g. owing to the very high thermal component
of Vela or to the closeness of J1418-6058 to an AGN). Table 2 provides details on such pulsars.\\
We consider an X-ray counterpart to be confirmed if:\\
- X-ray pulsation has been detected;\\
- X and Radio coordinates concide;\\
- X-ray source position has been validated through the blind-search algorithm developed by the Fermi collaboration \citep{Abd09c,Ray10}.\\
If none of these conditions apply, $\gamma$-ray pulsar is labelled as "0".\\
According to our classification scheme we have 14 type-0, 7 type-1 and 37 type-2 pulsars. In total
44 $\gamma$-ray neutron stars, 31 radio-lound and 13 radio-quiet have an X-ray counterpart.

Since the X-ray observation database is continuously growing, the results available in literature encompass only fractions of the X-ray data now available.
Moreover, they have been obtained with different versions of the standard analysis softwares or using different
techniques to account for the PWN contribution.
Thus, with the exception of the well-known and bright X-ray pulsars, 
such as Crab or Vela, we re-analyzed all the X-ray data pubblicly available
following an homogeneous procedure.
If only a small fraction of the data are publicly available, we quoted results from a literature search.

In order to assess the X-ray spectra of Fermi pulsars, we used photons with energy 0.3$<$E$<$10 keV collected
by Chandra/ACIS \citep{Gar03}, 
XMM-Newton \citep{Str01}, \citep{Tur01} and SWIFT/XRT \citep{Bur05}.  
We selected all the public observations (as of April 2010) that overlap the error box 
of Fermi pulsars or the Radio coordinates.\\
We neglected all Chandra/HRC observations owing to the lack of energy resolution of the instrument.
To analyze Chandra data, we used the Chandra Interactive Analysis of Observation software (CIAO version 4.1.2).
The Chandra point spread function depends on the off-axis angle: we used for all the point sources
an extraction area around the pulsar that cointains 90\% of the events. For instance, for on-axis sources we
selected all the photons inside a 2" radius circle, while we extracted photons
from the inner part of PWNs (excluding the 2" radius circle of the point source) in order to assess the nebular spectra: such
extended regions vary for pulsar to pulsar as a function of the nebula dimension and flux.\\
We analyzed all the XMM-Newton data (both from PN and MOS1/2 detectors)
with the XMM-Newton Science Analysis Software (SASv8.0). 
The raw observation data files (ODFs) were processed using standard pipeline tasks (epproc for PN, emproc for MOS data); we used only photons
with event pattern 0-4 for the PN detector and 0-12 for the MOS1/2 detectors.
When necessay, an accurate screening for soft proton flare events was done, following the prescription by \citet{DeL04}.\\
If, in addiction to XMM data, deep Chandra data were also available, we made an XMM spectrum of the entire PSR+PWN and used the Chandra higher resolution
in order to disentangle the two contributions. When only XMM-Newton data were available, the point source 
was analyzed by selecting all the photons inside a 20" radius circle while 
the whole PWN (with the exception of the 20" radius circle of the point source) was used in order to assess the nebular spectrum.\\
We analyzed all the SWIFT/XRT data with HEASOFT version 6.5 selecting all the photons inside a 20" radius circle.
If multiple data sets collected by the same instruments were found, spectra, response and effective area files
for each dataset were added by using the mathpha, addarf and addrmf HEASOFT tools.

All the spectra have been studied with XSPEC v.12 \citep{Arn96} choosing, whenever possible, the same
background regions for all the different observations of each pulsar. All the data were rebinned in order to 
have at least 25 counts per channel, as requested for the validity of $\chi^2$ statistic.\\
The XMM-Chandra cross calibration studies \citep{xmm} report only minor changes in flux
($<$10\%) between the two instruments. When both XMM and Chandra data were available, a constant has been introduced to
account for such uncertainty. Conversely, when the data were collected only by one instrument, a systematic error
was introduced.
All the PSRs and PWNs have been fitted with absorbed powerlaws; when statistically needed,
a blackbody component has been added to the pulsar spectrum. Since PWNs typically show a 
powerlaw spectrum with a photon index which steepens moderately as a function 
of the distance from the PSR \citep{Gae06}, we used only the inner part
of each PWN. 
Absorption along the line of sight has been obtained through the fitting procedure
but for the cases with very low statistic for which we used informations derived from
observations taken in different bands.

\subsection{X-ray Analysis}

For pulsars with a good X-ray coverage we carried out the following steps.\\
If only XMM-Newton public observations were available, we tried to take into account the PWN contribution.
First we searched the literature for any evidence of the presence of a PWN and, if nothing was found, we analyzed
the data to search for extended emission. If no evidence for the presence of a PWN was found, 
we used PN and MOS1/2 data in a simultaneous spectral fit.
On the other hand, if a PWN was present, its contribution was evaluated on a case by case basis. If the statistic was good enough, we studied
simultanously the inner region, containing both PSR and PWN, and the extended source region surrounding it.
The inner region data were described by two absorbed (PWN and PSR) powerlaws, while the outer one by a single (PWN) powerlaw.
The N$_H$ and the PWN photon index values were the same in the two (inner and outer) datasets.\\
When public Chandra data were available, we evaluated separately PSR and PWN (if any) in a similar way.

If both Chandra and XMM public data were available, we exploited Chandra space resolution to evaluate the PWN contribution by:\\
- obtaining two different spectra of the inner region (a), encompassing both PSR and PWN and of the outer region (b) encompassing only the PWN;\\
- extracting a total XMM spectrum (c) containing both PSR and PWN: this is the only way to take into account the XMM's larger PSF;\\
- fitting simultaneously a,b,c with two absorbed powerlaws and eventually (if statistically significant) an absorbed blackbody, using the same N$_H$; 
a constant moltiplicative was also introduced in order to account for a possible discrepancy between 
Chandra and XMM calibrations;\\
- forcing to zero the normalization(s) of the PSR model(s) in the Chandra outer region and freeing the other normalizations in the Chandra datasets; 
fixing the XMM PSR normalization(s) at the inner Chandra dataset one and the XMM PWN normalization at the inner+outer normalizations of the Chandra PWN.\\
Only for few well-known pulsars, or pulsars for which the dataset is not yet entirely public, we used results taken from the literature (see Table \ref{tab-2}). Where necessary,
we used XSPEC in order to obtain the flux in the 0.3-10 keV energy range and to evaluate the unabsorbed flux.\\

For pulsars with a confirmed counterpart but too few photons to discriminate the spectral shape, we evaluated an hypothetical unabsorbed
flux by assuming that a single powerlaw spectrum with a photon index of 2 to describe PSR+PWN. We also assumed that 
the PWN and PSR thermal contributions are 30\% of the entire source flux (a sort of mean value of all the
considered type 2 pulsars).
To evaluate the absorbing column, we need a distance value which can come either from
the radio dispersion or - for radio-quiet pulsars - from the following pseudo-distance reported in \citet{Saz10}:
$d=0.51\dot{E_{34}}^{1/4}/F_{\gamma,10}^{1/2}$ kpc \\
where $\dot{E}=\dot{E}_{34}\times10^{34}erg/s$ and $F_{\gamma}=F_{\gamma,10}\times10^{-10}erg/cm^2s$
and the beam correction factor f$_{\gamma}$ is assumed to be 1 \citep{Wat09} for all pulsars.\\
Then, the HEASARC WebTools (http://heasarc.gsfc.nasa.gov/docs/tools.html) was used to find the galactic column density (N$_H$) 
in the direction of the pulsar; with the distance information,
we could rescale the column density value of the pulsar.
We found the source count rate by using the XIMAGE task \citep{Gio92}.
Then, we used the WebPimms tool inside the WebTools package to evaluate the source unabsorbed flux.
Such a value has then to be corrected to account for the PWN and PSR thermal contributions.
We are aware that each pulsar can have a different
photon index, as well as thermal and PWN contributions so that we used these mean values only as a first
approximation. All the low-quality pulsars (type 1) will be treated separately and all the considerations in this paper
will be based only on high-quality objects (type 2).

For pulsars without a confirmed counterpart we evaluated the X-ray unabsorbed flux upper limit assuming
a single powerlaw spectrum with a photon index of 2 to describe PSR+PWN  and using a signal to noise of 3.\\
The column density has been evaluated as above.
Under the previous hypotheses, we used the signal-to-noise definition in order to compute the upper limit to the absorbed flux of the X-ray
counterpart. Next we used XSPEC to find the unabsorbed upper limit flux.

On the basis of our X-ray analysis we define a subsample of Fermi $\gamma$-ray pulsars for which we have, at once, 
reliable X-ray data (type 2 pulsars) and satisfactory distance estimates such parallax, radio dispersion
measurement, column density estimate, SNR association.
Such a subsample contains 24 radio emitting neutron stars and 5 radio quiet ones. The low number of radio quiet is to be
ascribed to lack of high quality X-ray data. Only one of the IBIS pulsars has a clear distance estimate.
Moreover, we have 4 additional radio-quiet pulsars with reliable X-ray data but without a satisfactory distance estimate.

In Tables \ref{tab-2}-\ref{tab-3} we reported the gamma-ray and X-ray parameters of the 54 Fermi first year pulsars. 
We also included the four hard X-ray pulsars taken from
the "4th IBIS/ISGRI soft gamma-ray survey catalog" \citep{Bir09}. We use $\dot{E}=4\pi^2I\dot{P}/P$, $\tau_C=P/2\dot{P}$ and 
$B_{lc}=3.3\times10^{19}(P\dot{P})^{1/2}\times(10km)/(R_{lc}^3)$, where $R_{lc}=cP/2\pi$, P is the pulsar spin period, $\dot{P}$ its derivative 
and the standard value for moment of inertia of the neutron star I=$10^{45}g/cm^2$ \citep[see e.g.][]{Ste10}.
Using the P and $\dot{P}$ values taken from \citet{Abd09,Saz10}, we computed
the values reported in Table \ref{tab-1}. Most of the distance values are taken from \citet{Abd09,Saz10} (see Table \ref{tab-1}).

\section{Discussion}

\subsection{Study of the X-ray luminosity}

The X-ray luminosity, L$_X$, 
is correlated with the pulsar spin-down luminosity $\dot{E}$. The scaling was firstly noted by \citet{Sew88} who used Einstein data of 22 pulsar
- most of them just upper limits - to derive a linear relation between log$F_{0.2-4 keV}^{X}$ and log$\dot{E}$. Later, \citet{Bec97} investigated 
a sample of 27 pulsars by using ROSAT, yielding the simple scaling L$_X^{0.1-2.4keV}\simeq10^{-3}\dot{E}$. The uncertainty due to soft X-ray absorption
translates into very high flux errors; moreover it was very hard to discriminate between the thermal and powerlaw spectral components.
A re-analysis was performed by \citet{Pos02}, who studied in the 2-10 keV band a sample of
39 pulsars observed by several X-ray telescopes. However, they could not separate
the PWN from the pulsar contribution. Moreover, they conservatively adopted, for most of the pulsars, an uncertainty of $40\%$ on the distance values.
A better comparison with our data can be done with the results by \citet{Kar08}, who recently used high-resolution Chandra
data in order to disentangle the PWN and pulsar fluxes. Focussing just on Chandra data, and rejecting XMM observations, they obtain a
poor spectral characterization which translates in high errors on fluxes. They also adopted an uncertainty of $40\%$ on the distance values for most pulsars.
Despite the big uncertainties, mainly due to poor distance estimates, all these datasets show that the L$_X$ versus $\dot{E}$ relation
is quite scattered. The high values of the $\chi^2_{red}$ seem to exclude a simple statistical effect.

We are now facing a different panorama, since our ability to evaluate pulsars' distances has improved \citep{Abd09,Saz10} and we are now much better in
discriminating pulsar emission from its nebula.
The use of XMM data makes it possible to build good quality spectra allowing to disentangle the non-thermal from the thermal 
contribution, when present. In particular, we can study the newly discovered radio-quiet
pulsar population and compare them with the "classical" radio-loud pulsars.
We investigate the relations between the X and $\gamma$ luminosities and pulsar parameters, making use of the data collected in Tables \ref{tab-1}-\ref{tab-2}-\ref{tab-3}.\\

Using the 29 Fermi type 2 pulsars with a clear distance estimate and with a well-constrained X-ray spectrum,
the weighted least square fit yields:
\begin{equation}
log_{10}L^X_{29}=(1.11_{-0.30}^{+0.21})+(1.04\pm0.09)log_{10}\dot{E}_{34}
\end{equation}
where $\dot{E}=\dot{E}_{34}\times10^{34}erg/s$ and $L_X=L^X_{29}\times10^{29}erg/s$. All the uncertains are at 90\% confidence level.
We can evaluate the goodness of this fit using the reduced chisquare value $\chi^2_{red}=3.7$;
a double linear fit does not significantly change the value of $\chi^2_{red}$.
A more precise way to evaluate the dispersion of the dataset around the fitted curve is the parameter:\\
$W^2=(1/n)\sum_{i=1->n}(y_{oss}^i-y_{fit}^i)^2$\\
where $y_{oss}^i$ is the actual i$^{th}$ value of the dataset (in our case $log_{10}L^X_{29}$) and $y_{fit}^i$ the expected one. A lesser spread in the dataset
translate into a lower value of $W^2$. We obtain $W^2=0.436$ for the $L_x-\dot{E}$ relationship.
Such high values of both $W^2$ and $\chi_{red}^2$ are an indication of an important scattering of the $L_X$ values around the fitted relation.\\
Our results are in agreement with \citet{Pos02,Kar08}.

\subsection{Study of the $\gamma$-ray luminosity}

The gamma-ray luminosity, L$_{\gamma}$, is correlated with the pulsar spin-down luminosity $\dot{E}$.
Such a trend is expected in many theorical models \citep[see e.g.][]{Zha04,Mus03}
and it's shortly discussed in the Fermi LAT catalog of gamma-ray pulsars \citep{Abd09}.

Selecting the same subsample of Fermi pulsar used in the previous chapter to assess the relation
between $L_{\gamma}$ and $\dot{E}$, we found that a linear fit:
\begin{equation}
log_{10}L_{32}^{\gamma}=(0.45_{-0.17}^{+0.50})+(0.88\pm0.07)log_{10}\dot{E}_{34}
\end{equation}
yields an high value of $\chi^2_{red}=4.2$.\\
Inspection of the distribution of residuals lead us to try a double-linear relationship, 
which yields:
\begin{mathletters}
\begin{eqnarray}
log_{10}L_{32}^{\gamma}=(2.45\pm0.76)+(0.20_{-0.31}^{+0.27})log_{10}\dot{E}_{34} & , & \dot{E}>E_{crit}\\
log_{10}L_{32}^{\gamma}=(0.52\pm0.18)+(1.43_{-0.23}^{+0.31})log_{10}\dot{E}_{34} & , & \dot{E}<E_{crit}
\end{eqnarray}
\end{mathletters}
with $E_{crit}=3.72_{-3.44}^{+3.55}\times10^{35}erg/s$ and $\chi^2_{red}=2.2$. An f-test shows
that the probability for a chance $\chi^2$ improvement is 0.00011.
Such a result is in agreement with the data reported in \citet{Abd09} for the entire dataset
of Fermi $\gamma$-ray pulsars.
Indeed, the $\chi^2_{red}$ obtained for the double linear fit is better than that obtained for the 
$L_X$-$\dot{E}$ relationship. We obtain $W^2=0.344$ for the double linear $L_{\gamma}-\dot{E}$ relationship. 
Both the $\chi^2_{red}$ and $W^2$ are in agreement with a little higher scatter in the $L_X-\dot{E}$ graph.
A difference between the X-ray and $\gamma$-ray emission geometries - that translates in different values of 
f$_{\gamma}$ and f$_X$ - could explain such a behaviour.

The existence of an $\dot{E}_{crit}$ has been posited from the theoretical point for different
pulsar emission models.
Revisiting the outer-gap model for pulsars with $\tau<10^7$ yrs and assuming initial conditions as well as
pulsars' birth rates, \citet{Zha04} found a sharp boundary, due to the saturation of the gap size, for $L_{\gamma}=\dot{E}$.
They obtain the following distribution of pulsars' $\gamma$-ray luminosities:
\begin{mathletters}
\begin{eqnarray}
log_{10}L_{\gamma}=log_{10}\dot{E}+const. & , &  \dot{E}<\dot{E}_{crit}\\
log_{10}L_{\gamma}\sim0.30log_{10}\dot{E}+const. & , & \dot{E}>\dot{E}_{crit}
\end{eqnarray}
\end{mathletters}
By assuming the fractional gap size from \citet{Zha97}, they obtain $\dot{E}_{crit}=1.5\times10^{34}P^{1/3}erg/s$.
While Equation 4 is similar to our double linear fit (Equation 3), the $\dot{E}_{crit}$ they obtain
seems to be lower than our best fit value.\\
On the other hand, in slot-gap models \citep{Mus03}, the break occurs
at about $10^{35}erg/s$, when the gap is limited by screening of the acceleration field by pairs.\\
We can see from Figure \ref{fig-2} that radio-quiet pulsars have higher luminosities than the radio-loud ones, for similar values of
$\dot{E}$. As in the $L_X-\dot{E}$ fit, we can't however discriminate between the two population due to the big errors stemming from distance estimate.

\subsection{Study of the $\gamma$-to-X ray luminosity ratio}
\label{sec-GX}

At variance with the X-ray and gamma-ray luminosities, the ratio between the X-ray and gamma-ray luminosities is indipendent
from pulsars' distances. This makes it possible to significatively reduce the error bars leading to more 
precise indications on the pulsars' emission mechanisms.\\
Figure \ref{fig-3} reports the histogram of the F$_{\gamma}$/F$_X$ values using only type 2 (high quality X-ray data) pulsars. 
The radio-loud pulsars have
 $<F_{\gamma}/F_X>\sim800$ while the radio-quiet population has $<F_{\gamma}/F_X>\sim4800$.
Applying the Kolmogoroff-Smirnov test to type 2 pulsars' $F_{\gamma}/F_X$ values we obtained that the chance for the two
datasets belong to the same population is 0.0016. By using all the pulsars with a confirmed X-ray counterpart (i.e. including
also type 1 objects) this probability increase to 0.00757.
We can conclude, with a 3$\sigma$ confidence level, that the radio-quiet and radio-loud datasets we used are somewhat different.

\subsubsection{A distance indipendent spread in F$_{\gamma}$/F$_X$}

\begin{figure}
\centering
\includegraphics[angle=0,scale=.60]{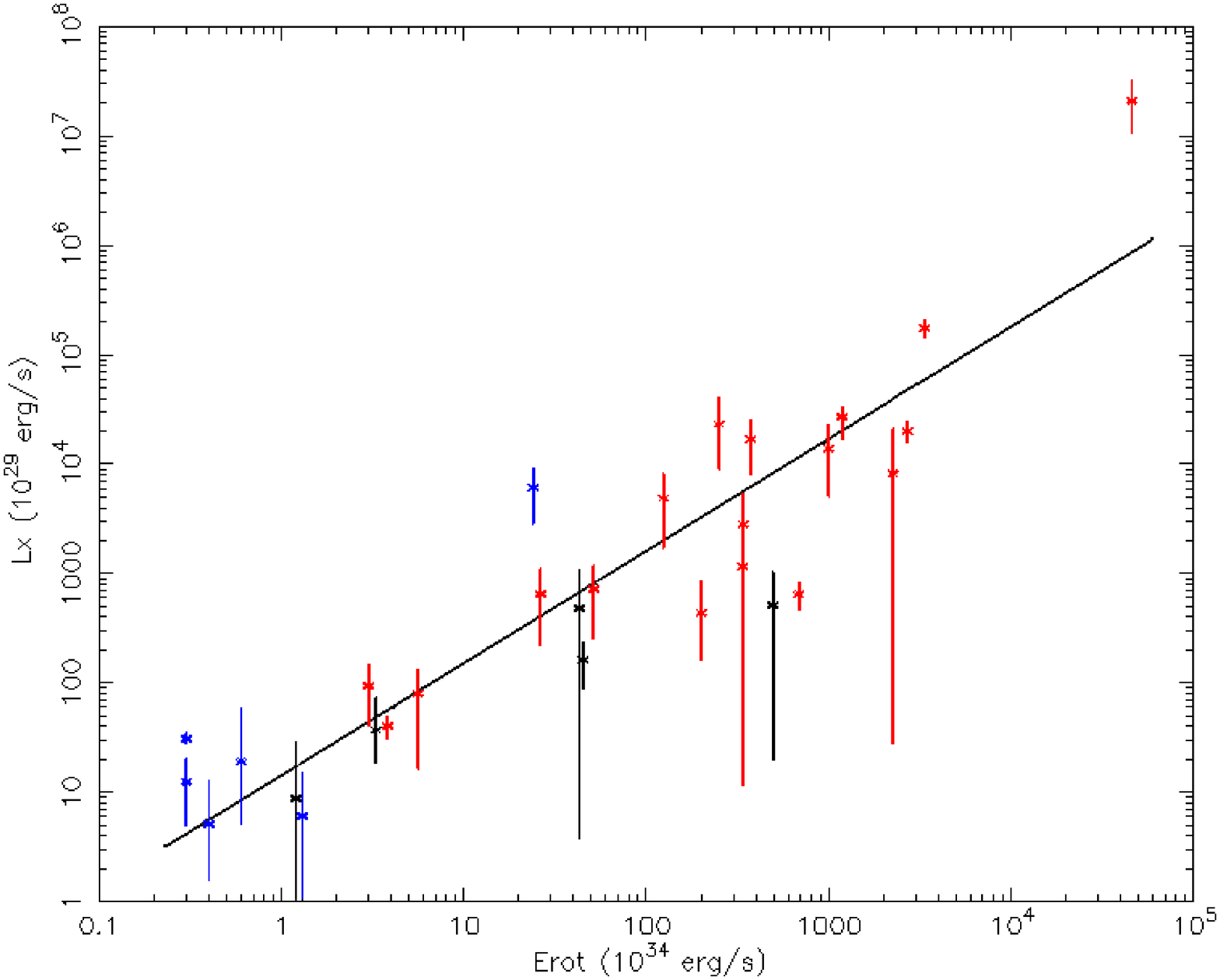}
\caption{$\dot{E}$-L$_X$ diagram for all pulsars classified as type 2 and with a clear distance estimation, assuming f$_X$=1 (see Equation 5). Black: radio-quiet pulsars; red: radio-loud pulsars; blue: millisecond pulsars. The linear best fit of the logs of the two quantities is shown. \label{fig-1}}
\end{figure}

\begin{figure}
\centering
\includegraphics[angle=0,scale=.60]{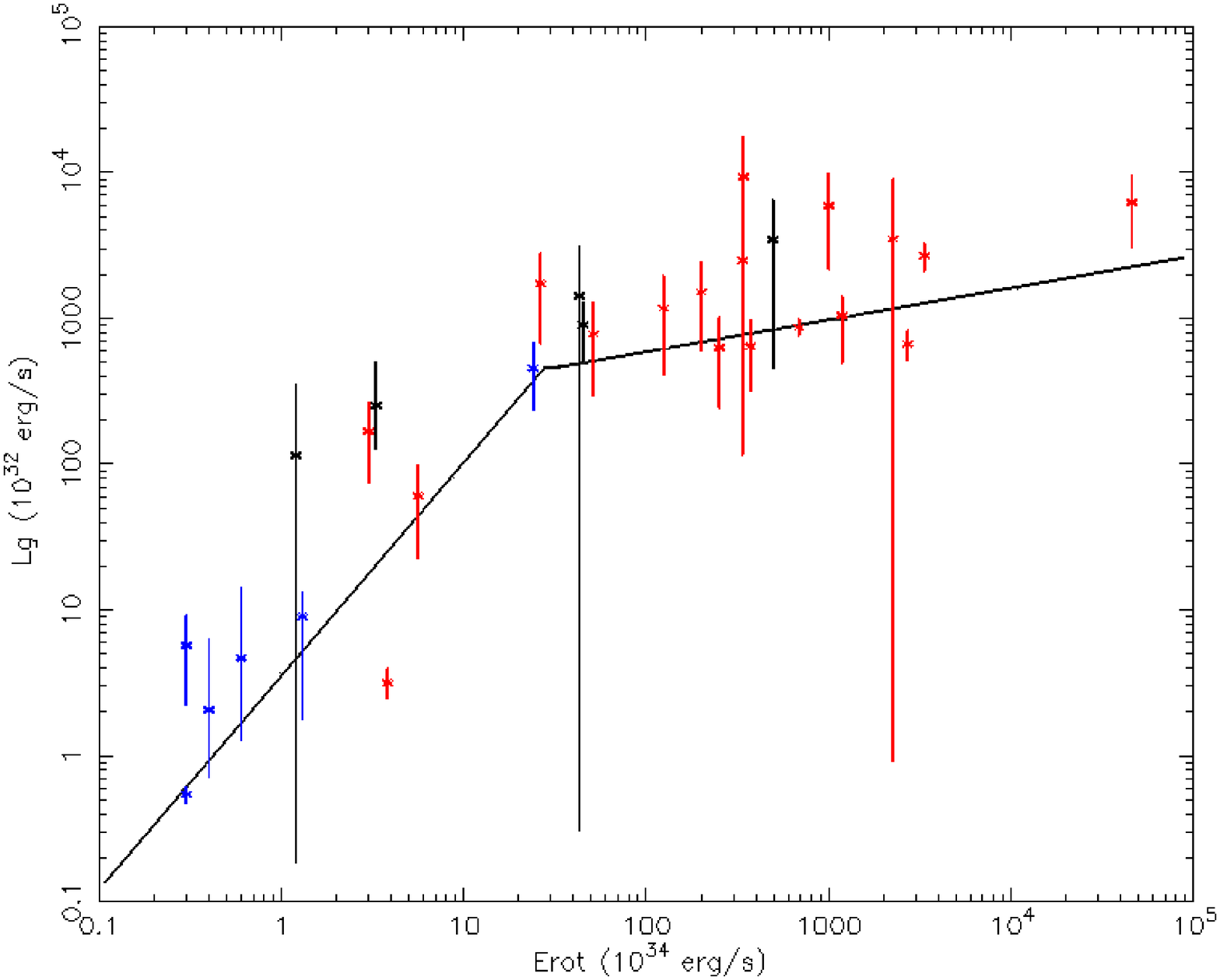}
\caption{$\dot{E}$-L$_{\gamma}$ diagram for all pulsars classified as type 2 and with a clear distance estimation, assuming f$_{\gamma}$=1 (see Equation 5). Black: radio-quiet pulsars; red: radio-loud pulsars; blue: millisecond pulsars. The double linear best fit of the logs of the two quantities is shown. \label{fig-2}}
\end{figure}

\begin{figure}
\centering
\includegraphics[angle=0,scale=.60]{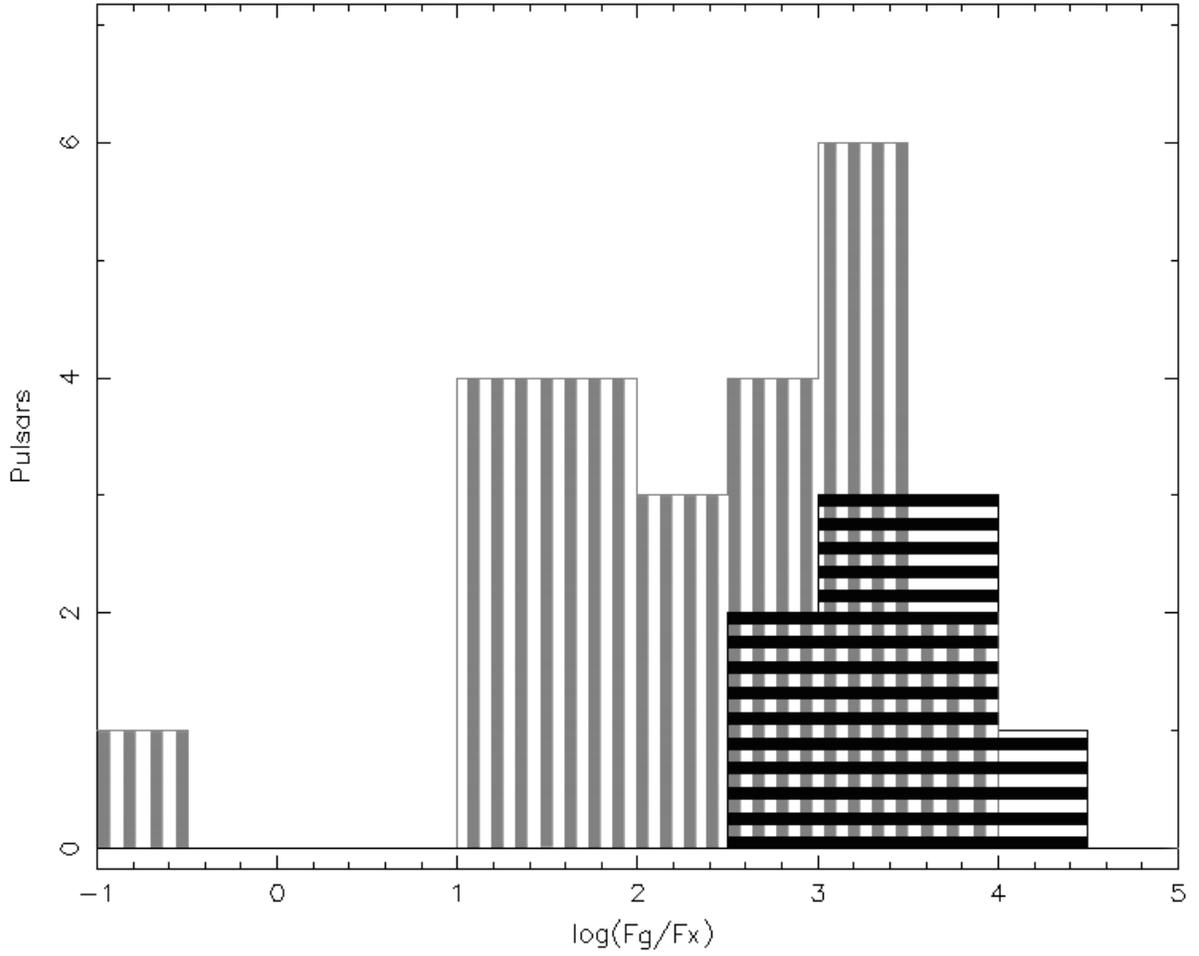}
\caption{log(F$_{\gamma}$/F$_X$) histogram. The step is 0.5; the radio-loud (and millisecond) pulsars are indicated in grey and the radio-quiet ones in black. Only high confidence pulsars (type 2) have been used for a total of 24 radio-loud and 9 radio-quiet pulsars. \label{fig-3}}
\end{figure}

\begin{figure}
\centering
\includegraphics[angle=0,scale=.60]{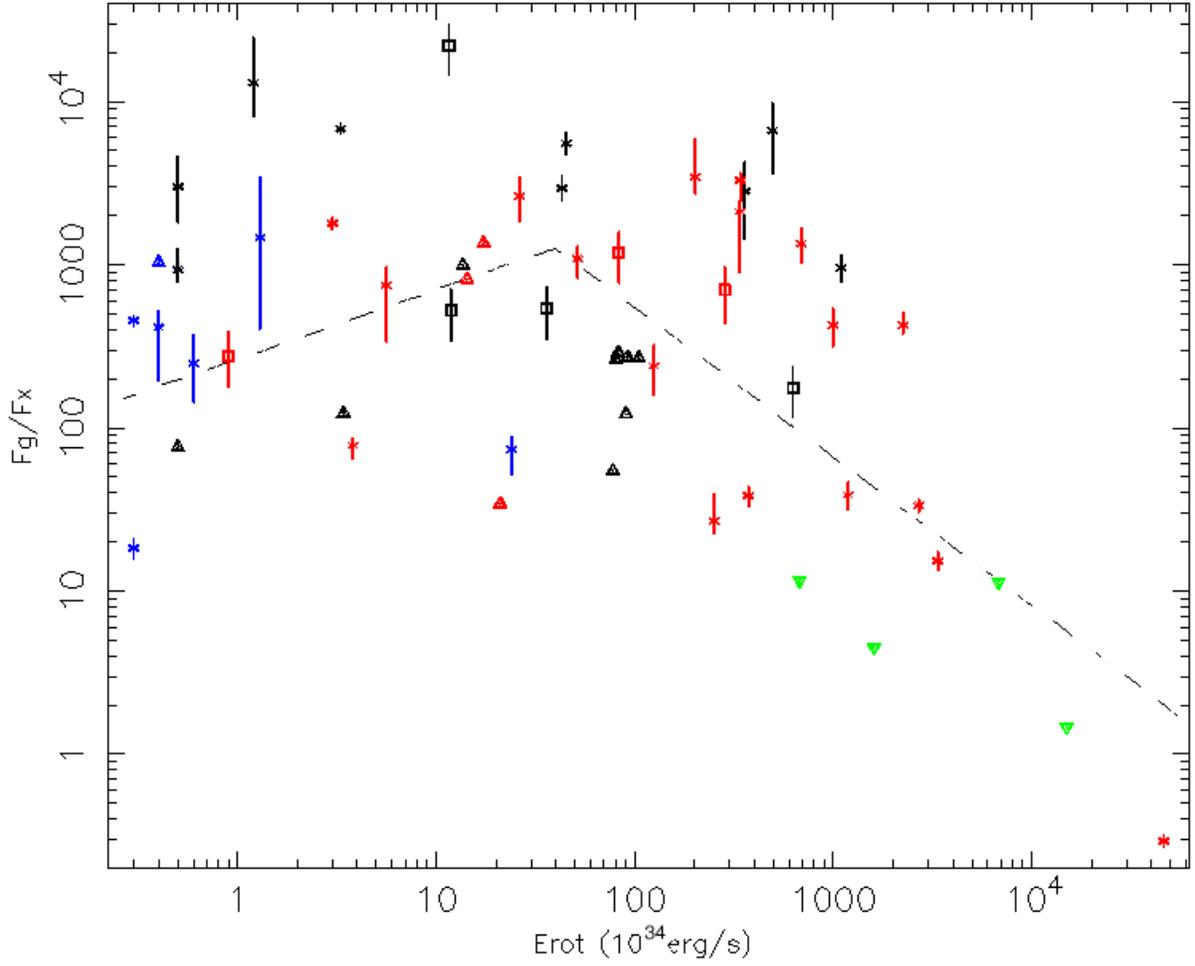}
\caption{$\dot{E}$-F$_{\gamma}$/F$_X$ diagram. Green: IBIS pulsars; black: radio-quiet pulsars; red: radio-loud pulsars; blue: millisecond pulsars. The triangles are upper and lower limits, the squares indicate pulsars with a type 1 X-ray spectrum (see Table \ref{tab-2}) and the stars pulsars with a high quality X-ray spectrum The dotted line is the combination of the best fitting functions obtained for Figure \ref{fig-1} and \ref{fig-2} with the geometrical correction factor set to 1 for both the X and $\gamma$-ray bands. \label{fig-4}}
\end{figure}

\begin{figure}
\centering
\includegraphics[angle=0,scale=.30]{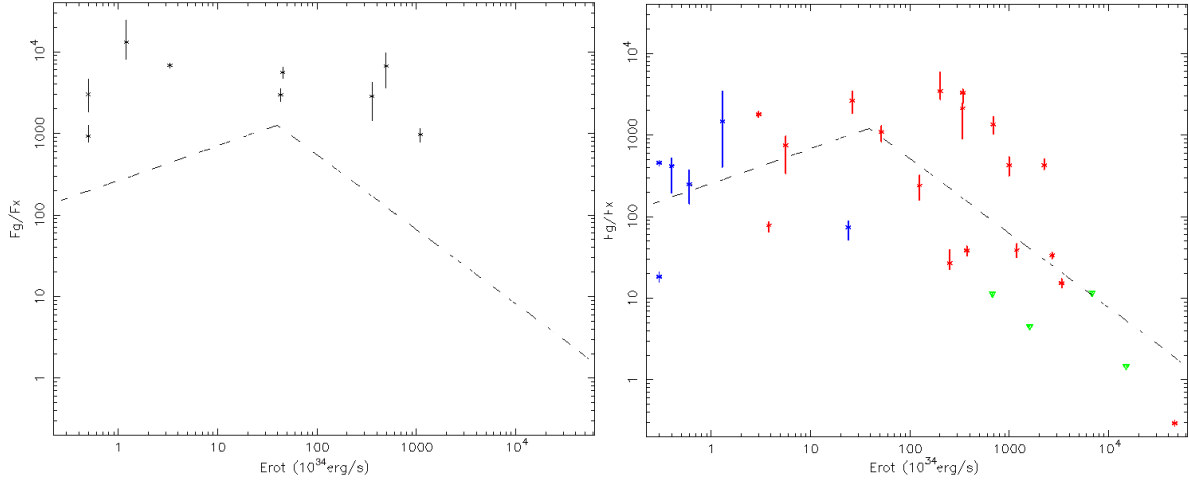}
\caption{$\dot{E}$-F$_{\gamma}$/F$_X$ diagram for high confidence pulsars only (type 2). Green: IBIS pulsars; black: radio-quiet pulsars; red: radio-loud pulsars; blue: millisecond pulsars. The triangles are upper limits. \label{fig-5}}
\end{figure}

Figure \ref{fig-4} shows F$_{\gamma}$/F$_X$ as a function of $\dot{E}$ for our entire sample of $\gamma$-ray emitting NSs
while in Fig \ref{fig-5} only the pulsar with "high quality" X-ray data have been selected.
Even neglecting the upper and lower limits (shown as triangles) as well as the low quality points (see Figure \ref{fig-5}), one immediatly notes
the scatter on the F$_{\gamma}$/F$_X$ parameter values for a given value of $\dot{E}$. Such an apparent spread
cannot obviously be ascribed to a low statistic. An inspection of Figure \ref{fig-4} makes it clear
that a linear fit cannot satisfactory describe the data. In a sense, this finding should not come as a surprise since Figure \ref{fig-4} is a combination
of Figures \ref{fig-1} and \ref{fig-2} and we have seen that figure \ref{fig-2} requires a double linear fit.
However, combining the results of our previous fits (Equations 1 and 3) we obtain the
dashed line in Figure \ref{fig-4}, clearly a very poor description of the data.
For $\dot{E}\sim<5\times10^{36}$ the F$_{\gamma}$/F$_X$ values scatter around a mean value of $\sim$1000 with a spread
of a factor about 100. For higher $\dot{E}$ the values of F$_{\gamma}$/F$_X$ seem to decrease drastically to an average value of $\sim$50,
reaching the Crab with F$_{\gamma}$/F$_X$$\sim0.1$.

The spread in the F$_{\gamma}$/F$_X$ values for pulsars with similar $\dot{E}$ is obviously unrelated to distance uncertainties. 
Such a scatter can be due to geometrical effects. For both X-ray and $\gamma$-ray energy bands:
\begin{equation}
L_{\gamma,X}=4\pi f_{\gamma,X}F_{obs}D^2
\end{equation}
where f$_X$ and f$_{\gamma}$ account for the X and $\gamma$ beaming geometries (which may or may not be related).
If the pulse profile observed along the line-of-sight at $\zeta$ (where $\zeta_E$ is the Earth line-of-sight) for a pulsar with
magnetic inclination $\alpha$ is $F(\alpha,\zeta,\phi)$, where $\phi$ is the pulse phase, than we can write:
\begin{equation}
f=f(\alpha,\zeta_E)=\frac{\int\int F(\alpha,\zeta,\phi)sin(\zeta)d\zeta d\phi}{2\int F(\alpha,\zeta_E,\phi)d\phi}
\end{equation}
where $f$ depends only from the viewing angle and the magnetic inclination of the pulsar. With an high
value of this correction coefficient, the emission is disfavoured. Obviously
F$_{\gamma}$/F$_X$=L$_{\gamma}$/L$_X\times$f$_X$/f$_{\gamma}$. Different $f_{\gamma}/f_X$ values
for different pulsars can explain the scattering seen in the F$_{\gamma}$/F$_X$-$\dot{E}$ relationship.\\
\citet{Wat09} assume a nearly uniform emission efficiency while
\citet{Zha04} compute a significant variation in the emission efficiency as a function of the geometry of pulsars.
In both cases, geometry plays an important role through magnetic field inclination as well as through viewing angle.\\
The very important scatter found for F$_{\gamma}$/F$_X$ values is obviously due to the different geometrical
configurations which determine the emission at different wavelength of each pulsar. While geometry is clearly playing
an equally important role in determining pulsar luminosities, the F$_{\gamma}$/F$_X$ plot makes its effect
easier to appreciate.

The dashed line in Figures \ref{fig-4} and \ref{fig-5} is the combination of the best fits of L$_{\gamma}$-$\dot{E}$ and 
L$_X$-$\dot{E}$ relationship, considering f$_{\gamma}$=1 and f$_X$=1 so that represent the hypothetical
value of F$_{\gamma}$/F$_X$ that each pulsar would have if f$_{\gamma}$=f$_X$: all the pulsars with a value of F$_{\gamma}$/F$_X$
below the line have f$_X<$f$_{\gamma}$.
We have seen in Section \ref{sec-GX} that the radio-quiet dataset shows an higher mean value of F$_{\gamma}$/F$_X$.
This is clearly visible in Figure \ref{fig-5} where all the radio-quiet points are above the expected values (dashed line)
so that all the radio-quiet pulsars should have f$_X>$f$_{\gamma}$. Moreover, the radio-quiet 
dataset shows a lower scatter with respect to the radio-loud one pointing to more uniform values of f$_{\gamma}$/f$_X$ 
for the radio-quiet pulsars. A similar viewing angle or a similar
magnetic inclination for all the radio-quiet pulsars could explain such a behaviour (see Equation 6).

\begin{figure}
\centering
\includegraphics[angle=0,scale=.30]{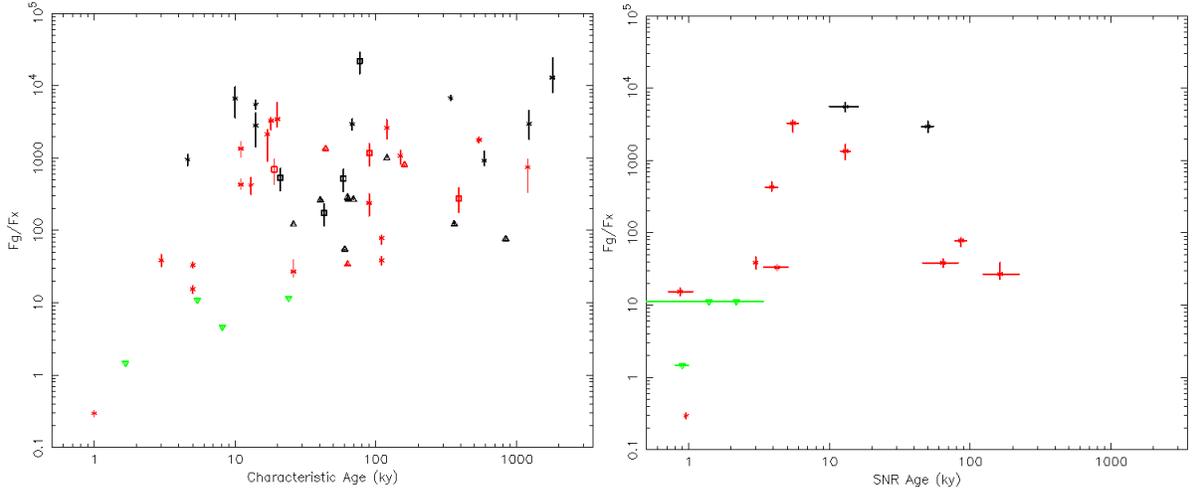}
\caption{Left: Characteristic Age-F$_{\gamma}$/F$_X$ diagram. Right: SNR Age-F$_{\gamma}$/F$_X$ diagram. Green: IBIS pulsars; black: radio-quiet pulsars; red: radio-loud pulsars. Triangles are upper limits, squares are pulsars with a type 1 X-ray spectrum while stars are pulsars with a type 2 X-ray spectrum. \label{fig-6}}
\end{figure}

Figure \ref{fig-6} shows the $F_{\gamma}/F_X$ behaviour as a function of the characteristic pulsar age. 
In view of the uncertainty of this parameter, we have also built a similar plot using "real" pulsar
age, as derived from the associated supernova remnants (see Figure 6).
Similarly to the $\dot{E}$ relationship, for $\tau<10^4$ years, $F_{\gamma}/F_X$ values increase with
age (both the characteristic and real ones), while for $t>10^4$years the behaviour becomes more complex.\\

\subsection{Study of the selection effects}

There are two main selections we have done in order to obtain our sample of pulsars with both good $\gamma$ and X-ray spectra (type 2).
First, the two populations of radio-quiet and radio-loud pulsars are unveiled with different techniques: using the same dataset, 
pulsars with known rotational ephemerides have a detection threshold
lower than pulsars found through blind period searches. In the First Fermi LAT pulsar catalog \citep{Abd09} the
faintest gamma-ray-selected pulsar has a flux $\sim$ 3$\times$ higher than the faintest radio-selected one.
Second, we chose only pulsars with a good X-ray coverage. Such a coverage depends on many factors (including the policy of
X-ray observatories) that cannot be modeled.\\
Our aim is to understand if these two selections influenced in different ways the two populations of pulsars we are studying:
if this was the case, the results obtained would have been distorced.
The $\gamma$-ray selection is discussed at length in the Fermi LAT pulsar catalog \citep{Abd09}. Since the radio-quiet population
has obviously a detection threshold higher than the radio-loud, we could avoid such bias by selecting all the pulsars with a flux higher than
the radio-quiet detection threshold (6$\times10^{-8}$ph/cm$^2$s).
Only five radio-loud type 2 pulsars are excluded (J0437-4715, J0613+1036, J0751+1807, J2043+2740 and J2124-3358)
with F$_{\gamma}$/F$_X$ values ranging from 87 to 1464. We performed our analysis on such a reduced sample
and the results don't change significatively.\\
We can, therefore, exclude the presence of an important bias due to the $\gamma$-ray selection on type 2 pulsars.

In order to roughly evaluate the selection affecting the X-ray observations, we used the method developed by \citep{Sch68} to compare the current
radio-quiet and -loud samples' spatial distributions, following the method also used in \citet{Abd09}.
For each object with an available distance estimate, we computed the maximum distance
still allowing detection from $D_{max}=D_{est}(F_{\gamma}/F_{min})^{1/2}$, where $D_{est}$ comes from Table 1, 
the photon flux and $F_{min}$ are taken from \citet{Abd09,Saz10}. 
We limited $D_{max}$ to 15 kpc, and compared $V$, the volume enclosed within the estimated source distance, to that
enclosed within the maximum distance,$V_{max}$, for a galactic disk with radius 10 kpc and thickness 1 kpc (as in \citet{Abd09}). The inferred values of $<V/V_{max}>$
are 0.462, 0.424, 0.443 and 0.516 for the entire gamma-ray pulsars' dataset, the radio-quiet pulsars, millisecond pulsars and
the radio-loud pulsars. These are quite close to the expected value of 0.5 even if $<V/V_{max}>^{rq}$ is lower than $<V/V_{max}>^{rl}$.
If we use only type 2 pulsars we obtain 0.395, 0.335, 0.462 and 0.419.
These lower values of $<V/V_{max}>$ indicate that we have a good X-ray coverage only for close-by - or very bright - pulsars,
not a surprising result. By using the X-ray-counterpart dataset, both the radio-loud and -quiet $<V/V_{max}>$ values appear lower of about 0.1:
this seems to indicate that we used the same selection criteria for the two population and we minimized 
the selection effects in the histogram of Figure \ref{fig-3}.\\
We can conclude that the $\gamma$-ray selection introduced no changes in the two populations,
while the X-ray selection excluded objects both faint and/or far away; any distortion, if present,
is not overwhelming.\\
Only if deep future X-ray observations centered on radio-quiet Fermi pulsars will fail to unveil lower
values of F$_{\gamma}$/F$_x$, it will be possible to be sure that 
radio-quiet pulsars have a different geometry (or a different emission mechanism) than radio-loud ones.

\section{Conclusions}

The discovery of a number of radio-quiet pulsars comparable to that of radio-loud ones
together with the study of their X-ray counterparts made it possible, for the first time,
to address their behaviour using a distance independent parameters such as the ratio of their fluxes at X and gamma ray wavelengths.\\
First, we reproduced the well known relationship between the neutron stars luminosities and their rotational energy losses. Next, selecting only the Fermi pulsars with good X-ray data,  we computed  the ratio between the gamma and X-ray fluxes and studied its dependence on the overall rotational energy loss as well as on the neutron star age.\\
Much to our surprise, the distance independent F$_{\gamma}$/F$_X$ values computed for pulsars of similar age and energetic differ by up to 3 orders of magnitude, pointing to important (yet poorly understood) differences both in position and height of the regions emitting at X and $\gamma$-ray wavelengths within the pulsars’ magnetospheres. Selection effects cannot account for the spread in the F$_{\gamma}$/F$_X$ relationship and any 
further distortion, if present, is not overwhelming.\\
In spite of the highly scattered values, a decreasing trend is seen when considering young and energetic pulsars.
Moreover, radio quiet pulsars are characterized by higher values of the F$_{\gamma}$/F$_X$ parameter  ($<F_{\gamma}/F_X>_{rl}\sim800$ and $<F_{\gamma}/F_X>_{rq}\sim4800$) so that a KS test points to a chance of 0.0016 for them to belong to the same population as the radio loud  ones. While it would be hard to believe that radio loud and radio quiet pulsars belong to two different neutron star populations, the KS test probably points to different geometrical configurations (possibly coupled with viewing angles) that characterize radio loud and radio quiet pulsars. Indeed 
the radio-quiet population we analyzed is less scattered than the radio-loud one, pointing to a more uniform viewing or 
magnetic geometry of radio-quiet pulsars.

Our work is just a starting point, based on the first harvest of gamma-ray pulsars. The observational panorama will quickly evolve. The gamma-ray pulsar list will certainly grow and this will trigger more X-ray observations, improving both in quantity and in quality the database of the neutron stars detected in X and $\gamma$-rays to be used to compute our multiwavelength, distance independent parameter. However, to fully exploit the information packed in the F$_{\gamma}$/F$_X$ a complete 3D modelling of pulsar magnetosphere is needed to account for the different locations and heights of the emitting regions at work at different energies. Such modelling could provide the clue to account for the spread we have observed for the ratios between $\gamma$ and X-ray fluxes as well as for the systematically higher values measured for radio-quiet pulsars.   

\acknowledgments

XMM data analysis is supported by contracts ASI-INAF I/088/06/0 and NASA NIPR NNG10PL01I30. Chandra data analysis is supported by 
INAF-ASI contract n.I/023/05/0.\\
It is a pleasure to thank Niel Gehrels for granting SWIFT observations of the newly disocvered Fermi pulsar. We also thank Pablo
Saz Parkinson and Andrea Belfiore for the collaboration in searching for X-ray counterparts of radio-quiet pulsars.

\begin{footnotesize}

\end{footnotesize}

\setlength{\LTleft}{-1pt}
\begin{footnotesize}
\begin{landscape}
\begin{center}
\begin{longtable}{cccccccccc}
\tableline\tableline
PSR Name & P$^a$ & $\dot{P}^a$ & $\tau_c$ & $\tau_{snr}^b$ & B$_{lc}$ & d$^a$ & $\dot{E}$ & Type$^e$ & PWN$^f$ \\
 & (ms) & $(10^{-15})$ & (ky) & (ky) & (kG) & (kpc) & ($10^{34}$erg/s) & & \\
\tableline\endhead
J0007+7303 & 316 & 361 & 14 & 13 & 3.1 & 1.4$\pm$0.3 & 45.2 & g & Y\\
J0030+0451 & 4.9 & $10^{-5}$ & 7.7$\times10^{6}$ & - & 17.8 & 0.30$\pm$0.09 & 0.3 & m & N\\
J0205+6449 & 65.7 & 194 & 5 & 4.25$\pm$0.85 & 115.9 & 2.9$\pm$0.3 & 2700 & r & Y\\
J0218+4232 & 2.3 & 7.7$\times10^{-5}$ & 5$\times10^{5}$ & - & 313.1 & 3.25$\pm$0.75 & 24 & m & N\\
J0248+6021 & 217 & 55.1 & 63 & - & 3.1 & 5.5$\pm3.5$ & 21 & r & ?\\
J0357+32  & 444 & 12 & 590 & - & 0.2 & 0.5$^c$ & 0.5 & g & Y\\
J0437-4715 & 5.8 & 1.4$\times10^{-5}$ & 6.6$\times10^{6}$ & - & 13.7 & 0.1563$\pm$0.0013 & 0.3 & m & N\\
J0534+2200 & 33.1 & 423 & 1.0 & 0.955 & 950 & 2.0$\pm$0.5 & 46100 & r & Y\\
J0540-6919 & 50.5 & 480 & 1.67 & 0.9$\pm$0.1 & 364 & 50$^d$ & 15000 & i & Y\\
J0613-0200 & 3.1 & 9$\times10^{-6}$ & 5.3$\times10^{6}$ & - & 54.3 & 0.48$^{+0.19}_{-0.11}$ & 1.3 & m & N\\
J0631+1036 & 288 & 105 & 44 & - & 2.1 & 2.185$\pm$1.440 & 17.3 & r & ?\\
J0633+0632 & 297 & 79.5 & 59 & - & 1.7 & 1.1$^c$ & 11.9 & g & ?\\
J0633+1746 & 237 & 11 & 340 & - & 1.1 & 0.250$_{-0.062}^{+0.12}$ & 3.3 & g & N\\
J0659+1414 & 385 & 55 & 110 & 86$\pm$8 & 0.7 & 0.288$_{-0.027}^{+0.033}$ & 3.8 & r & N\\
J0742-2822 & 167 & 16.8 & 160 & - & 3.3 & $2.07_{-1.07}^{+1.38}$ & 14.3 & r & ?\\
J0751+1807 & 3.5 & 6.2$\times10^{-6}$ & 8$\times10^{6}$ & - & 32.3 & 0.6$_{-0.2}^{+0.6}$ & 0.6 & m & N\\
J0835-4510 & 89.3 & 124 & 11 & 13$\pm$1 & 43.4 & 0.287$_{-0.017}^{+0.019}$ & 688 & r & Y\\
J1023-5746 & 111 & 384 & 4.6 & - & 44 & 2.4$^c$ & 1095 & g & ?\\
J1028-5819 & 91.4 & 16.1 & 90 & - & 14.6 & 2.33$\pm$0.70 & 83.2 & r & Y\\
J1044-5737 & 139 & 54.6 & 40.3 & - & 9.5 & 1.5$^c$ & 80.3 & g & ?\\
J1048-5832 & 124 & 96.3 & 20 & - & 16.8 & 2.71$\pm$0.81 & 201 & r & Y\\
J1057-5226 & 197 & 5.8 & 540 & - & 1.3 & 0.72$\pm$0.20 & 3.0 & r & N\\
J1124-5916 & 135 & 747 & 3 & 2.99$\pm$0.06 & 37.3 & 4.8$_{-1.2}^{+0.7}$ & 1190 & r & Y\\
J1413-6205 & 110 & 27.7 & 62.9 & - & 12.3 & 1.4$^c$ & 82.7 & g & ?\\
J1418-6058 & 111 & 170 & 10 & - & 29.4 & 3.5$\pm$1.5 & 495 & g & Y\\
J1420-6048 & 68.2 & 83.2 & 13 & - & 69.1 & 5.6$\pm$1.7 & 1000 & r & N\\
J1429-5911 & 116 & 30.5 & 60.2 & - & 11.3 & 1.6$^c$ & 77.5 & g & ?\\
J1459-60 & 103 & 25.5 & 64 & - & 13.6 & 1.5$^c$ & 91.9 & g & ?\\
J1509-5850 & 88.9 & 9.2 & 150 & - & 11.8 & 2.6$\pm$0.8 & 51.5 & r & N\\
J1614-2230 & 3.2 & 4$\times10^{-6}$ &  1.2$\times10^{6}$ & - & 33.7 & 1.27$\pm$0.39 & 0.5 & m & N\\
J1617-5055 & 69 & 135 & 8.13 & - & 86.6 & 6.5$\pm$0.4$^d$ & 1600 & i & Y\\
J1709-4429 & 102 & 93 & 18 & 5.5$\pm$0.5 & 26.4 & 2.5$\pm$1.1 & 341 & r & Y\\
J1718-3825 & 74.7 & 13.2 & 90 & - & 21.9 & 3.82$\pm$1.15 & 125 & r & Y\\
J1732-31 & 197 & 26.1 & 120 & - & 2.7 & 0.6$^c$ & 13.6 & g & ?\\
J1741-2054 & 414 & 16.9 & 390 & - & 0.3 & 0.38$\pm$0.11 & 0.9 & r & ?\\
J1744-1134 & 4.1 & 7$\times10^{-6}$ & 9$\times10^{6}$ & - & 24 & 0.357$_{-0.035}^{+0.043}$ & 0.4 & m & N\\
J1747-2958 & 98.8 & 61.3 & 26 & 163$_{-39}^{+60}$ & 23.5 & 2.0$\pm$0.6 & 251 & r & Y\\
J1809-2332 & 147 & 34.4 & 68 & 50$\pm$5 & 6.5 & 1.7$\pm$1.0 & 43 & g & Y\\
J1811-1926 & 62 & 41 & 24 & 2.18$\pm$1.22 & 64 & 7$\pm$2$^d$ & 678 & i & Y\\ 
J1813-1246 & 48.1 & 17.6 & 43 & - & 76.2 & 2.0$^c$ & 626 & g & ?\\
J1813-1749 & 44.7 & 150 & 5.4 & 1.3925$\pm$1.1075 & 272 & 4.70$\pm$0.47$^d$ & 680 & i & Y\\
J1826-1256 & 110 & 121 & 14 & - & 25.2 & 1.2$^c$ & 358 & g & Y\\
J1833-1034 & 61.9 & 202 & 5 & 0.87$_{-0.15}^{+0.20}$ & 137.3 & 4.7$\pm$0.4 & 3370 & r & Y\\
J1836+5925 & 173 & 1.5 & 1800 & - & 0.9 & 0.4$_{-0.15}^{+0.4 d}$ & 1.2 & g & N\\
J1846+0919 & 226 & 9.93 & 360 & - & 1.2 & 1.2$^c$ & 3.4 & g & ?\\
J1907+06 & 107 & 87.3 & 19 & - & 23.2 & 1.3$^c$ & 284 & r & ?\\
J1952+3252 & 39.5 & 5.8 & 110 & 64.0$\pm$18 & 71.6 & 2.0$\pm$0.5 & 374 & r & Y\\
J1954+2836 & 92.7 & 21.2 & 69.5 & - & 16.4 & 1.7$^c$ & 105 & g & ?\\
J1957+5036 & 375 & 7.08 & 838 & - & 0.3 & 0.9$^c$ & 0.5 & g & ?\\
J1958+2841 & 290 & 222 & 21 & - & 3.0 & 1.4$^c$ & 35.8 & g & ?\\
J2021+3651 & 104 & 95.6 & 17 & - & 26 & 2.1$_{-1.0}^{+2.1}$ & 338 & r & Y\\
J2021+4026 & 265 & 54.8 & 77 & - & 1.9 & 1.5$\pm$0.45 & 11.6 & g & ?\\
J2032+4127 & 143 & 19.6 & 120 & - & 5.3 & 3.60$\pm$1.08 & 26.3 & r & ?\\
J2043+2740 & 96.1 & 1.3 & 1200 & - & 3.6 & 1.80$\pm$0.54 & 5.6 & r & N\\
J2055+25 & 320 & 4.08 & 1227 & - & 0.3 & 0.4$^c$ & 0.5 & g & ?\\
J2124-3358 & 4.9 & 1.2$\times10^{-5}$ & 6$\times10^{5}$ & - & 18.8 & 0.25$_{-0.08}^{+0.25}$ & 0.4 & m & N\\
J2229+6114 & 51.6 & 78.3 & 11 & 3.90$\pm$0.39 & 134.5 & 3.65$\pm$2.85 & 2250 & r & Y\\
J2238+59 & 163 & 98.6 & 26 & - & 8.6 & 2.1$^c$ & 90.3 & g & ?\\
\tableline
\caption{\\
a : P, $\dot{P}$ and most of the values of the distance are taken from \citet{Abd09}, \citet{Saz10}.\\
b : Age derived from the associated SNR. Respectively taken from \citet{Sla04a}, \citet{Got07a}, \citet{Rud08a}, \citet{Hwa01a}, \citet{Tho03a}, \citet{Gor74a}, \citet{Win09a}, \citet{Bock02a}, \citet{Hal09a}, \citet{Rob08a}, \citet{Tam08a}, \citet{Bro05a}, \citet{Bie08a}, \citet{Mig08a}, \citet{Kot06}.\\
c : These distances are taken from \citet{Saz10} and are obtained under the absumption of a beam correction factor f$_{\gamma}$ = 1 for the gamma-ray emission cone of all pulsars. In this way one obtains:\\
$d=0.51\dot{E_{34}}^{1/4}/F_{\gamma,10}^{1/2}$ kpc\\
where $\dot{E}=\dot{E}_{34}\times10^{34}erg/s$ and $F_{\gamma}=F_{\gamma,10}\times10^{-10}erg/cm^2s$.
See also \citet{Saz10}.\\
d : Respectively taken from \citet{Cam08}, \citet{Kas98}, \citet{Kas01}, \citet{Got09}, \citet{Hal07}.\\
e : g = radio-quiet pulsars ; r = radio-loud pulsars ; m = millisecond pulsars ; i = pulsars detected by INTEGRAL/IBIS but not yet by Fermi (see \citet{Bir09}).\\
f : Only bright PWNs have been considered (with F$^{pwn}_x$ $>$ 1/5 F$^{psr}_x$). The presence or the absence of a bright PWN has been valued by re-analyzing the X-ray data (except for the X-ray analyses taken from literature, see the following table).\\
\label{tab-1}}
\end{longtable}

\begin{longtable}{cccccccccc}
\tableline\tableline
PSR Name & X$^a$ & Inst$^b$ & F$_X^{nt}$ & F$_X^{tot}$ & N$_H$ & $\gamma_X$ & kT & R$_{BB}$ & Eff$_X^d$\\
 & & & ($10^{-13}erg/cm^{2}s$) & ($10^{-13}erg/cm^{2}s$) & ($10^{20}cm^{-3}$) & & (keV) & (km) & \\
\tableline\endhead
J0007+7303 & 2 & X/C & 0.686$\pm$0.100 & 0.841$\pm$0.098 & 16.6$_{-7.6}^{+8.9}$ & 1.30$\pm$0.18 & 0.102$_{-0.018}^{+0.032}$ & 0.64$_{-0.20}^{+0.88}$ & 2.84$\times10^{-5}$\\
J0030+0451 & 2 & X & 1.16$\pm$0.02 & 2.8$\pm$0.1 & 0.244$_{-0.244}^{+7.470}$ & 2.8$_{-0.4}^{+0.5}$ & 0.194$_{-0.021}^{+0.015}$ & 0.6$_{-0.1}^{+0.225}$ & 3.32$\times10^{-4}$\\
J0205+6449 & 2 & C & 19.9$\pm$0.5 & 19.9$\pm$0.5 & 40.2$\pm$0.11 & 1.82$\pm$0.03 & - & - & 5.92$\times10^{-5}$\\
J0218+4232 & 2 & L$^f$ & 4.87$_{-1.28}^{+0.57}$ & 4.87$_{-1.28}^{+0.57}$ & 7.6$\pm$4.3 & 1.19$\pm$0.12 & - & - & 2.05$\times10^{-3}$\\
J0248+6021 & 0 & S & $<$9.00 & $<$9.00 & 80$^c$ & 2 & - & - & -\\
J0357+32 & 2 & C & 0.64$_{-0.06}^{+0.09}$ & 0.64$_{-0.06}^{+0.09}$ & 8.0$\pm$4.0 & 2.53$\pm$0.25 & - & - & -\\
J0437-4715 & 2 & X/C & 10.1$_{-0.6}^{0.8}$ & 14.3$_{-0.7}^{0.9}$ & 4.4 $\pm$ 1.7 & 3.17 $\pm$ 0.13 & 0.228$_{-0.003}^{+0.006}$ & 0.060$_{-0.008}^{+0.009}$ & 8.23$\times10^{-5}$\\
J0534+2200 & 2 & L$^f$ & 44300$\pm$1000 & 44300$\pm$1000 & 34.5$\pm$0.2 & 1.63$\pm$0.09 & - & - & 3.67$\times10^{-3}$\\
J0540-6919 & 2 & L$^f$ & 568$\pm$6 & 568$\pm$6 & 37$\pm$1 & 1.98$\pm$0.02 & - & - & 1.13$\times10^{-1}$\\
J0613-0200 & 2* & X & 0.221$_{-0.158}^{+0.297}$ & 0.221$_{-0.158}^{+0.297}$ & 1$^e$ & 2.7$\pm$0.4 & - & - & 3.74$\times10^{-5}$\\
J0631+1036 & 0 & X & $<$0.225 & $<$0.225 & 20$^c$ & 2 & - & - & -\\
J0633+0632 & 1 & S & 1.53$\pm$0.51 & 1.53$\pm$0.51 & 20$^c$ & 2 & - & - & -\\
J0633+1746 & 2 & L$^f$ & 4.97$_{-0.27}^{+0.09}$ & 12.6$_{-0.7}^{+0.2}$ & 1.07$^e$ & 1.7$\pm$0.1 & 0.190$\pm$0.030 & 0.04$\pm$0.01 & 8.99$\times10^{-5}$\\
J0659+1414 & 2 & L$^f$ & 4.06$_{-0.59}^{+0.03}$ & 168$_{-24}^{+1}$ & 4.3$\pm$0.2 & 2.1$\pm$0.3 & 0.125$\pm$0.003 & 1.80$\pm$0.15 & 8.46$\times10^{-5}$\\
J0742-2822 & 0 & X & $<$0.225 & $<$0.225 & 20$^c$ & 2 & - & - & -\\
J0751+1807 & 2 & L$^f$ & 0.44$_{-0.13}^{+0.18}$ & 0.44$_{-0.13}^{+0.18}$ & 4$^e$ & 1.59$\pm$0.30 & - & - & 2.52$\times10^{-4}$\\
J0835-4510 & 2* & L$^f$ & 65.1$\pm$15.7 & 281$\pm$67 & 2.2$\pm$0.5 & 2.7$\pm$0.6 & 0.129$\pm$0.007 & 2.5$\pm$0.3 & 9.78$\times10^{-6}$\\
J1023-5746 & 2* & C & 1.61$\pm$0.27 & 1.61$\pm$0.27 & 115$_{-41}^{+47}$ & 1.15$_{-0.22}^{+0.24}$ & - & - & -\\
J1028-5819 & 1 & S & 1.5$\pm$0.5 & 1.5$\pm$0.5 & 50$^c$ & 2 & - & - & -\\
J1044-5737 & 0 & S & $<$3.93 & $<$3.93 & 50$^c$ & 2 & - & - & -\\
J1048-5832 & 2* & C+X & 0.50$_{-0.10}^{+0.35}$ & 0.50$_{-0.10}^{+0.35}$ & $90_{-20}^{+40}$ & 2.4$\pm$0.5 & - & - & 1.74$\times10^{-5}$\\
J1057-5226 & 2 & C+X & 1.51$_{-0.13}^{+0.02}$ & 24.5$_{-2.5}^{+0.3}$ & 2.7$\pm$0.2 & 1.7$\pm$0.1 & 0.179$\pm$0.006 & 0.46$\pm$0.06 & 2.49$\times10^{-4}$\\
J1124-5916 & 2 & C & 9.78$_{-1.03}^{+1.18}$ & 10.90$_{-1.26}^{+1.32}$ & 30.0$_{-4.8}^{+2.8}$ & 1.54$_{-0.17}^{+0.09}$ & 0.426$_{-0.018}^{+0.034}$ & 0.274$_{-0.077}^{+0.089}$ & 2.27$\times10^{-4}$\\
J1413-6205 & 0 & S & $<$4.9 & $<$4.9 & 40$^c$ & 2 & - & - & -\\
J1418-6058 & 2 & C+X & 0.353$\pm$0.154 & 0.353$\pm$0.154 & 233$_{-106}^{+134}$ & 1.85$_{-0.56}^{+0.83}$ & - & - & 1.05$\times10^{-5}$\\
J1420-6048 & 2* & X & 1.6$\pm$0.7 & 1.6$\pm$0.7 & 202$_{-106}^{+161}$ & 0.84$_{-0.37}^{+0.55}$ & - & - & 1.11$\times10^{-4}$\\
J1429-5911 & 0 & S & $<$16.9 & $<$16.9 & 80$^c$ & 2 & - & - & -\\
J1459-60 & 0 & S & $<$3.93 & $<$3.93 & 100$^c$ & 2 & - & - & -\\
J1509-5850 & 2 & C+X & 0.891$_{-0.186}^{+0.132}$ & 0.891$_{-0.186}^{+0.132}$ & 80$^e$ & 1.31$\pm$0.15 & - & - & 1.12$\times10^{-4}$\\
J1614-2230 & 0 & C+X & $<$0.286 & 0.286$_{-0.086}^{+0.015}$ & 2.9$_{-2.9}^{+4.3}$ & 2 & 0.236$\pm$0.024 & 0.92$_{-0.35}^{+0.73}$ & -\\
J1617-5055 & 2 & L$^f$ & 64.2$\pm$0.3 & 64.2$\pm$0.3 & 345$\pm$14 & 1.14$\pm$0.06 & - & - & 2.03$\times10^{-3}$\\
J1709-4429 & 2 & C+X & 3.78$_{-0.94}^{+0.37}$ & 9.04$_{-2.25}^{+0.87}$ & 45.6$_{-2.9}^{+4.4}$ & 1.88$\pm$0.21 & 0.166$\pm$0.012 & 4.3$_{-0.86}^{+1.72}$ & 6.62$\times10^{-5}$\\
J1718-3825 & 2 & X & 2.80$\pm$0.67 & 2.80$\pm$0.67 & 70$^e$ & 1.4$\pm$0.2 & - & - & 3.12$\times10^{-4}$\\
J1732-31 & 0 & S & $<$2.42 & $<$2.42 & 50$^c$ & 2 & - & - & -\\
J1741-2054$^{g}$ & 1 & S & 4.64$_{-1.63}^{+1.84}$ & 4.64$_{-1.63}^{+1.84}$ & 0$^e$ & 2.10$_{-0.28}^{+0.50}$ & - & - & 9.93$\times10^{-4}$\\
J1744-1134 & 0 & C & $<$0.272 & 0.272$\pm$0.020 & 12$_{-12}^{+42}$ & 2 & 0.272$_{-0.098}^{+0.094}$ & 0.132$_{-0.120}^{+1.600}$ & -\\
J1747-2958 & 2* & C+X & 48.7$_{-6.0}^{+21.3}$ & 48.7$_{-6.0}^{+21.3}$ & 256$_{-6}^{+9}$ & 1.51$_{-0.44}^{+0.12}$ & - & - & 7.41$\times10^{-4}$\\
J1809-2332 & 2 & C+X & 1.40$_{-0.23}^{+0.25}$ & 3.14$_{-0.53}^{+0.57}$ & 61$_{-8}^{+9}$ & 1.85$_{-0.36}^{+1.89}$ & 0.190$\pm$0.025 & 1.54$_{-0.44}^{+1.26}$ & 8.98$\times10^{-5}$\\
J1811-1926 & 2 & C & 26.6$_{-3.7}^{+2.3}$ & 26.6$_{-3.7}^{+2.3}$ & 175$_{-12}^{+11}$ & 0.91$_{-0.08}^{+0.09}$ & - & - & 1.18$\times10^{-3}$\\
J1813-1246 & 1 & S & 9.675$\pm$3.225 & 9.675$\pm$3.225 & 100$^c$ & 2 & - & - & 1.13$\times10^{-3}$\\
J1813-1749 & 2 & C & 24.4$\pm$11.5 & 24.4$\pm$11.5 & 840$_{-373}^{+433}$ & 1.3$\pm$0.3 & - & - & -\\
J1826-1256 & 2 & C & 1.18$\pm$0.58 & 1.18$\pm$0.58 & 100$^e$ & 0.63$_{-0.63}^{+0.90}$ & - & - & -\\
J1833-1034 & 2 & X+C & 66.3$\pm$2.0 & 66.3$\pm$2.0 & 230$^e$ & 1.51$\pm$0.07 & - & - & 4.15$\times10^{-4}$\\
J1836+5925 & 2 & X+C & 0.459$_{-0.174}^{+0.403}$ & 0.570$_{-0.216}^{+0.500}$ & 0$_{-0}^{+0.792}$ & 1.56$_{-0.73}^{+0.51}$ & 0.056$_{-0.009}^{+0.012}$ & 4.47$_{-1.31}^{+3.03}$ & 5.84$\times10^{-5}$\\
J1846+0919 & 0 & S & $<$2.92 & $<$2.92 & 20$^c$ & 2 & - & - & -\\
J1907+06 & 1 & C & 3.93$\pm$1.45 & 3.93$\pm$1.45 & 398$_{-375}^{+468}$ & 3.16$_{-2.28}^{+2.76}$ & - & - & -\\ 
J1952+3252 & 2 & L$^f$ & 35.0$\pm$4.4 & 38.0$\pm$3.0 & 30$\pm$1 & 1.63$_{-0.05}^{+0.03}$ & 0.13$\pm$0.02 & 2.2$_{-0.8}^{+1.4}$ & 3.57$\times10^{-4}$\\
J1954+2836 & 0 & S & $<$3.65 & $<$3.65 & 50$^c$ & 2 & - & - & -\\
J1957+5036 & 0 & S & $<$2.98 & $<$2.98 & 10$^c$ & 2 & - & - & -\\
J1958+2841 & 1 & S & 1.57$\pm$0.53 & 1.57$\pm$0.53 & 40$^c$ & 2 & - & - & -\\
J2021+3651 & 2 & C+X & 2.21$_{-1.27}^{+0.35}$ & 6.01$_{-3.44}^{+0.96}$ & 65.5$\pm$6.0 & 2$\pm$0.5 & 0.140$_{-0.018}^{+0.023}$ & 4.94$\pm$1.40 & 2.75$\times10^{-5}$\\
J2021+4026 & 1 & C & 0.443$\pm$0.148 & 0.443$\pm$0.148 & 40$^c$ & 2 & - & - & 1.03$\times10^{-4}$\\
J2032+4127 & 2* & C+X & 0.423$\pm$0.118 & 0.423$\pm$0.118 & 38.7$_{-38.7}^{+75.6}$ & 1.87$_{-0.76}^{+0.96}$ & - & - & 1.99$\times10^{-4}$\\
J2043+2740 & 2 & X & 0.208$_{-0.208}^{+0.480}$ & 0.208$_{-1.08}^{+0.48}$ & 0$_{-0}^{+20}$ & 3.1$\pm$0.4 & - & - & 1.44$\times10^{-4}$\\
J2055+25 & 2 & X & 0.382$_{-0.148}^{+0.197}$ & 0.382$_{-0.148}^{+0.197}$ & 7.3$_{-7.3}^{+10.4}$ & 2.2$_{-0.6}^{+0.5}$ & - & - & 1.79$\times10^{-3}$\\
J2124-3358 & 2 & X & 0.668$_{-0.344}^{+0.150}$ & 0.959$_{-0.494}^{+0.216}$ & 2.76$_{-2.76}^{+4.87}$ & 2.89$_{-0.35}^{+0.45}$ & 0.268$_{-0.032}^{+0.034}$ & 0.019$_{-0.009}^{+0.012}$ & 1.25$\times10^{-4}$\\
J2229+6114 & 2 & C+X & 51.3$_{-5.8}^{+9.3}$ & 51.3$_{-5.8}^{+9.3}$ & 30$_{-4}^{+9}$ & 1.01$_{-0.12}^{+0.06}$ & - & - & 2.90$\times10^{-4}$\\
J2238+59 & 0 & S & $<$4.49 & $<$4.49 & 70$^c$ & 2 & - & - & -\\
\tableline
\caption{X-ray spectra of the pulsars. The fluxes are unabsorbed and here the non-thermal and total fluxes are shown. The model used is an absorbed powerlaw plus blackbody, where statistically necessary. The only exceptions are PSR J0437-4715 (double PC plus powerlaw), J0633+1746 and J0659+1414 (double BB plus powerlaw): here only the most relevant thermal component is reported. All the errors are at a 90\% confidence level.\\
a : This parameter shows the confidency of the X-ray spectrum of each pulsar, based on the available X-ray data. An asterisk mark the pulsars for which ad ad-hoc analysis was necessary. See section ~\ref{xdata}.\\
b : C = Chandra/ACIS ; X = XMM/PN+MOS ; S = SWIFT/XRT ; L = literature. Only public data have been used (at December 2010).\\
c : here, the column density has been fixed by using the galactic value in the pulsar direction obtained by Webtools (http://heasarc.gsfc.nasa.gov/docs/tools.html) and scaling it for the distance (see Table \ref{tab-1}).\\
d : the beam correction factor f$_X$ is assumed to be 1, which can result in an efficiency $>$ 1. See \citet{Wat09}. Here the errors are not reported.\\
e : The statistic is very low so that it was necessary to freeze the column density parameter; the values have been evaluated by using WebTools (http://heasarc.gsfc.nasa.gov/docs/tools.html).\\
f : Respectively taken from \citet{Web04a}, \citet{Kar08}, \citet{Cam08}, \citet{DeL05}, \citet{DeL05}, \citet{Web04b}, \citet{Mor04}, \citet{Kar09}, \citet{Li05}.\\
g : The spectrum is well fitted also by a single blackbody.\\
\label{tab-2}}
\end{longtable}

\begin{longtable}{cccccccc}
\tableline\tableline
PSR Name & F$_R$ & F$_{\gamma}$ & $\gamma_{\gamma}$ & Cutoff$_G$ & Eff$_{\gamma}^a$ & F$_{\gamma}$/F$_{X nt}$ & F$_{\gamma}$/F$_{X tot}$\\
 & (mJy) & ($10^{-10}erg/cm^{2}s$)& & (GeV) & & & \\ 
\tableline\endhead
J0007+7303 & $<$0.006$^c$ & 3.82$\pm$0.11 & 1.38$\pm$0.04 & 4.6$\pm$0.4 & 0.2 & 5570$\pm$827 & 4544$\pm$546\\
J0030+0451 & 0.6 & 0.527$\pm$0.035 & 1.22$\pm$0.16 & 1.8$\pm$0.4 & 0.17 & 454$\pm$31 & 188$\pm$14\\
J0205+6449 & 0.04 & 0.665$\pm$0.054 & 2.09$\pm$0.14 & 3.5$\pm$1.4 & 0.0025 & 33.4$\pm$2.9 & 33.4$\pm$2.9\\
J0218+4232 & 0.9 & 0.362$\pm$0.053 & 2.02$\pm$0.23 & 5.1$\pm$4.2 & 0.2 & 74.3$_{-22.4}^{+13.9}$ & 74.3$_{-22.4}^{+13.9}$\\
J0248+6021 & 9 & 0.308$\pm$0.058 & 1.15$\pm$0.49 & 1.4$\pm$0.6 & 0.735 & $>$34.2 & $>$34.2\\
J0357+32 & $<$0.043$^c$ & 0.639$\pm$0.037 & 1.29$\pm$0.18 & 0.9$\pm$0.2 & 5.23 & 1000$_{-150}^{+200}$ & 1000$_{-150}^{+200}$\\
J0437-4715 & 140 & 0.186$\pm$0.022 & 1.74$\pm$0.32 & 1.3$\pm$0.7 & 0.02 & 18.4$\pm$2.5 & 13.0$\pm$1.7\\
J0534+2200 & 14 & 13.07$\pm$1.12 & 1.97$\pm$0.06 & 5.8$\pm$1.2 & 0.001 & 0.295$\pm$0.026 & 0.295$\pm$0.026\\
J0540-6919 & 0.024 & $<$0.833 & 2 & - & - & $<$1.47 & $<$1.47\\
J0613-0200 & 1.4 & 0.324$\pm$0.035 & 1.38$\pm$0.24 & 2.7$\pm$1.0 & 0.07 & $1464_{-1059}^{+1974}$ & $1464_{-1059}^{+1974}$\\
J0631+1036 & 0.8 & 0.304$\pm$0.051 & 1.38$\pm$0.35 & 3.6$\pm$1.8 & 0.14 & $>$1350 & $>$1350\\
J0633+0632 & $<$0.003$^c$ & 0.801$\pm$0.064 & 1.29$\pm$0.18 & 2.2$\pm$0.6 & 1.4 & 524$\pm$179 & 524$\pm$179\\
J0633+1746 & $<$1 & 33.85$\pm$0.29 & 1.08$\pm$0.02 & 1.90$\pm$0.05 & 0.78 & 6812$_{-375}^{+136}$ & 2687$_{-151}^{+48}$\\
J0659+1414 & 3.7 & 0.317$\pm$0.030 & 2.37$\pm$0.42 & 0.7$\pm$0.5 & 0.01 & 78.1$_{-13.6}^{+7.5}$ & 1.89$_{-0.32}^{+0.18}$\\
J0742-2822 & 15 & 0.183$\pm$0.035 & 1.76$\pm$0.40 & 2.0$\pm$1.4 & 0.07 & $>$812 & $>$812\\
J0751+1807 & 3.2 & 0.109$\pm$0.032 & 1.56$\pm$0.58 & 3.0$\pm$4.3 & 0.08 & 248$_{-103}^{+125}$ & 248$_{-103}^{+125}$\\
J0835-4510 & 1100 & 88.06$\pm$0.45 & 1.57$\pm$0.01 & 3.2$\pm$0.06 & 0.01 & 1353$\pm$326 & 313$\pm$75\\
J1023-5746 & $<$0.031 & 1.55$\pm$0.10 & 1.47$\pm$0.14 & 1.6$\pm$0.3 & 0.12 & 963$\pm$173 & 963$\pm$173\\
J1028-5819 & 0.36 & 1.77$\pm$0.12 & 1.25$\pm$0.17 & 1.9$\pm$0.5 & 0.14 & 1182$\pm$403 & 1182$\pm$403\\
J1044-5737 & $<$0.021 & 1.03$\pm$0.07 & 1.60$\pm$0.12 & 2.5$\pm$0.5 & 0.45 & $>$262 & $>$262\\
J1048-5832 & 6.5 & 1.73$\pm$0.11 & 1.31$\pm$0.15 & 2.0$\pm$0.4 & 0.08 & 3451$_{-725}^{+2426}$ & 3451$_{-725}^{+2426}$\\
J1057-5226 & 11 & 2.72$\pm$0.08 & 1.06$\pm$0.08 & 1.3$\pm$0.1 & 0.56 & 1804$_{-164}^{+59}$ & 111$_{-12}^{+4}$\\
J1124-5916 & 0.08 & 0.380$\pm$0.058 & 1.43$\pm$0.33 & 1.7$\pm$0.7 & 0.01 & 38.9$\pm$7.4 & 34.9$\pm$6.7\\
J1413-6205 & $<$0.025 & 1.29$\pm$0.10 & 1.32.$\pm$0.16 & 2.6$\pm$0.6 & 0.43 & $>$287 & $>$287\\
J1418-6058 & $<$0.03$^c$ & 2.36$\pm$0.32 & 1.32$\pm$0.20 & 1.9$\pm$0.4 & 0.08 & 6672$\pm$3049 & 6672$\pm$3049\\
J1420-6048 & 0.9 & 1.59$\pm$0.29 & 1.73$\pm$0.20 & 2.7$\pm$1.0 & 0.06 & 426$\pm$112 & 426$\pm$112\\
J1429-5911 & $<$0.022 & 0.926$\pm$0.081 & 1.93$\pm$0.14 & 3.3$\pm$1.0 & 0.45 & $>$55.0 & $>$55.0\\
J1459-60 & $<$0.038$^c$ & 1.06$\pm$0.10 & 1.83$\pm$0.20 & 2.7$\pm$1.1 & 0.52 & $>$269 & $>$269\\
J1509-5850 & 0.15 & 0.969$\pm$0.101 & 1.36$\pm$0.23 & 3.5$\pm$1.1 & 0.15 & 1088$_{-254}^{+197}$ & 1088$_{-254}^{+197}$\\
J1614-2230 & - & 0.274$\pm$0.042 & 1.34$\pm$0.36 & 2.4$\pm$1.0 & 1.03 & $>$958 & 958$_{-434}^{+196}$\\
J1617-5055 & - & $<$0.3 & 2 & - & - & $<$4.5 & $<$4.5\\
J1709-4429 & 7.3 & 12.42$\pm$0.22 & 1.70$\pm$0.03 & 4.9$\pm$0.4 & 0.33 & 3285$_{-819}^{+327}$ & 1374$_{-343}^{+134}$\\
J1718-3825 & 1.3 & 0.673$\pm$0.160 & 1.26$\pm$0.62 & 1.3$\pm$0.6 & 0.09 & 240$\pm$81 & 240$\pm$81\\
J1732-31 & $<$0.008$^c$ & 2.42$\pm$0.12 & 1.27$\pm$0.12 & 2.2$\pm$0.3 & 1.33 & $>$1000 & $>$1000\\
J1741-2054 & 0.156$^c$ & 1.28$\pm$0.07 & 1.39$\pm$0.14 & 1.2$\pm$0.2 & 0.24 & 277$_{-98}^{+111}$ & 277$_{-98}^{+111}$\\
J1744-1134 & 3 & 0.280$\pm$0.046 & 1.02$\pm$0.59 & 0.7$\pm$0.4 & 0.1 & $>$1030 & 1030$\pm$187\\
J1747-2958 & 0.25 & 1.31$\pm$0.14 & 1.11$\pm$0.28 & 1.0$\pm$0.2 & 0.02 & 26.9$_{-4.3}^{+12.1}$ & 26.9$_{-4.3}^{+12.1}$\\
J1809-2332 & $<$0.026$^c$ & 4.13$\pm$0.13 & 1.52$\pm$0.06 & 2.9$\pm$0.3 & 0.33 & 2951$_{-494}^{+535}$ & 1316$_{-226}^{+242}$\\
J1811-1926 & - & $<$0.3 & 2 & - & - & $<$11.25 & $<$11.25\\
J1813-1246 & $<$0.028$^c$ & 1.69$\pm$0.11 & 1.83$\pm$0.12 & 2.9$\pm$0.8 & 0.20 & 175$\pm$59 & 175$\pm$59\\
J1813-1749 & - & $<$0.3 & 2 & - & - & $<$11.25 & $<$11.25\\
J1826-1256 & $<$0.044$^c$ & 3.34$\pm$0.15 & 1.49$\pm$0.09 & 2.4$\pm$0.3 & 20.7 & 2834$\pm$1398 & 2834$\pm$1398\\
J1833-1034 & 0.07 & 1.02$\pm$0.12 & 2.24$\pm$0.15 & 7.7$\pm$4.8 & 0.01 & 15.3$\pm$1.9 & 15.3$\pm$1.9\\
J1836+5925 & $<$0.01$^c$ & 6.00$\pm$0.11 & 1.35$\pm$0.03 & 2.3$\pm$0.1 & 2 & 13065$_{-4958}^{+11474}$ & 10520$_{-3991}^{+9231}$\\
J1846+0919 & $<$0.004 & 0.358$\pm$0.035 & 1.60$\pm$0.19 & 4.1$\pm$1.5 & 2.1 & $>$123 & $>$123\\
J1907+06 & 0.0034$^b$ & 2.75$\pm$0.13 & 1.84$\pm$0.08 & 4.6$\pm$1.0 & 0.30 & 700$\pm$260 & 700$\pm$260\\
J1952+3252 & 1 & 1.34$\pm$0.07 & 1.75$\pm$0.10 & 4.5$\pm$1.2 & 0.02 & 38.3$\pm$5.3 & 35.2$\pm$3.4\\
J1954+2836 & $<$0.004 & 0.975$\pm$0.068 & 1.55$\pm$0.14 & 2.9$\pm$0.7 & 0.39 & $>$267 & $>$267\\
J1957+5036 & $<$0.025 & 0.227$\pm$0.020 & 1.12$\pm$0.28 & 0.9$\pm$0.2 & 5.6 & $>$76.2 & $>$76.2\\
J1958+2841 & $<$0.005$^c$ & 0.846$\pm$0.069 & 0.77$\pm$0.26 & 1.2$\pm$0.2 & 1.6 & 539$\pm$187 & 539$\pm$187\\
J2021+3651 & 0.1 & 4.70$\pm$0.15 & 1.65$\pm$0.06 & 2.6$\pm$0.3 & 0.07 & 2129$_{-1225}^{+343}$ & 783$_{-449}^{+127}$\\
J2021+4026 & $<$0.011$^c$ & 9.77$\pm$0.18 & 1.79$\pm$0.03 & 3.0$\pm$0.2 & 2.2 & 22061$\pm$7381 & 22061$\pm$7381\\
J2032+4127 & 0.05$^c$ & 1.12$\pm$0.12 & 0.68$\pm$0.38 & 2.1$\pm$0.6 & 0.64 & 2636$\pm$790 & 2636$\pm$790\\
J2043+2740 & 7 & 0.155$\pm$0.027 & 1.07$\pm$0.55 & 0.8$\pm$0.3 & 0.09 & 747$_{-409}^{+217}$ & 747$_{-409}^{+217}$\\
J2055+25 & $<$0.106 & 1.15$\pm$0.07 & 0.71$\pm$0.19 & 1.0$\pm$0.2 & 5.4 & 3010$_{-1181}^{+1563}$ & 3010$_{-1181}^{+1563}$\\
J2124-3358 & 1.6 & 0.276$\pm$0.035 & 1.05$\pm$0.28 & 2.7$\pm$1.0 & 0.05 & 413$_{-219}^{+107}$ & 288$_{-153}^{+74}$\\
J2229+6114 & 0.25 & 2.20$\pm$0.08 & 1.74$\pm$0.07 & 3.0$\pm$0.5 & 0.025 & 42.9$_{-5.1}^{+7.9}$ & 42.9$_{-5.1}^{+7.9}$\\
J2238+59 & $<$0.007$^c$ & 0.545$\pm$0.059 & 1.00$\pm$0.36 & 1.0$\pm$0.3 & 0.52 & $>$121 & $>$121\\
\tableline
\caption{$\gamma$-ray spectra of the pulsars. A broken powerlaw spectral shape is assumed for all the pulsars and the values are taken from \citet{Abd09,Saz10}. The gamma-ray flux is above 100 GeV. The 4 sources with an upper limit flux are taken from \citet{Bir09} (see section ~\ref{gdata}). The radio flux densities (at 1400MHz) are taken from \citet{Abd09,Saz10}. All the errors are at a 90\% confidence level.\\
a : f${_\gamma}$ is assumed to be 1, which can result in an efficiency $>$ 1. See \citet{Wat09}. Here the errors are not reported.\\
b : taken from \citet{Abd10e}.\\
c : taken from \citet{Ray10}
\label{tab-3}}
\end{longtable}
\end{center}
\end{landscape}
\end{footnotesize}

\end{document}